\documentclass[aps,twocolumn,showpacs,floats,prl]{revtex4}

\usepackage{graphicx}
\usepackage{dcolumn}
\usepackage{bm}
\usepackage{color}
\usepackage{subfigure}

\usepackage[normalem]{ulem}

\begin{document}

\author{
  A. Macia$^{\rm a}$, G. E. Astrakharchik$^{\rm a}$,
  F. Mazzanti$^{\rm a}$, S. Giorgini$^{\rm b}$, and J. Boronat$^{\rm a}$
}

\affiliation{
 $^{\rm a}$ Departament de F\'{i}sica i Enginyeria Nuclear,
  Universitat Polit\`{e}cnica de Catalunya, Campus Nord B4-B5,
  E-08034, Barcelona, Spain \\
$^{\rm b}$ Dipartimento di Fisica, Universit\`a di Trento and INO-CNR
  BEC Center, 38123 Povo, Trento, Italy
}

\title{Single-particle vs. pair superfluidity in a bilayer system of   dipolar bosons}

\begin{abstract}
We consider the ground state of a bilayer system of dipolar bosons, where dipoles are oriented by an external field in the direction perpendicular to the parallel planes.  Quantum Monte Carlo methods are used to calculate the ground-state energy, the one-body and two-body density matrix, and the superfluid response as a function of the separation between layers. We find that by decreasing the interlayer distance for fixed value of the strength of the dipolar interaction, the system undergoes a quantum phase transition from a single-particle to a pair superfluid. The single-particle superfluid is characterized by a finite value of both the atomic condensate and the super-counterfluid density. The pair superfluid phase is found to be stable against formation of many-body cluster states and features a gap in the spectrum of elementary excitations.
\end{abstract}

\maketitle

The study of quantum degenerate gases of dipolar particles has become
in recent years one of the most active areas of experimental and
theoretical research in the field of ultracold atoms~\cite{Review1,
  Review2}. The realization of systems featuring strong dipolar
interactions opens prospects for investigating new and highly
interesting many-body effects which arise from the anisotropic and
long-range nature of the interatomic force. An example is the quest
for $p$-wave superfluidity in a two-dimensional (2D) Fermi gas where
dipoles are aligned by an external field at an angle formed with the
plane of confinement larger than some critical
value~\cite{Bruun08,Sieberer11}. Another example involves fermionic
dipoles in a bilayer geometry that allows for interlayer pairing of
particles and displays superfluidity of pairs which, depending on the
interlayer distance, ranges from a Bardeen-Cooper-Schrieffer (BCS)
type to a Bose-Einstein condensate (BEC) of tightly bound
dimers~\cite{Pikovski10, Matveeva14}. This latter system shares many
analogies with the electron-hole bilayers realized in semiconductor
coupled quantum wells~\cite{Sivan92, Seamons09} as well as
graphene~\cite{Gorbachev12}, where excitonic superfluidity is
predicted to occur~\cite{Lozovik75, Perali13} even though a clear
experimental observation is still lacking.

In this Letter, we investigate two-dimensional bilayers of bosonic dipoles,
where the dipoles are oriented perpendicularly to the parallel planes which
provide the 2D confinement. Tunneling between layers is assumed to be
negligible due to the high potential barrier separating the planes. If one
neglects short-range forces, in-plane interactions are purely repulsive and
behave as $1/r^3$ in terms of the interparticle distance. On the contrary,
out-of-plane interactions are attractive at short distance and might induce pairing
between particles in the two layers.

In contrast to the fermionic counterpart, the bosonic system displays a
quantum phase transition, as a function of the interlayer attraction, from
a single-particle to a pair superfluid state (see Fig.~\ref{fig1}).  In the
case of a tight-binding model of hard-core bosons on a lattice, the phase
diagram at zero temperature has been investigated using mean
field~\cite{Trefzger09} and quantum Monte Carlo (QMC)
methods~\cite{Pupillo13} and was found to include exotic phases around
half filling such as the checkerboard solid and the pair supersolid. Here
we study a translationally invariant system without optical lattices
in the planes and we perform simulations in the continuum by means of the
diffusion Monte Carlo (DMC) technique. The ground-state energy is
calculated as a function of the distance between the two layers and the
emergence of off-diagonal long-range order (ODLRO) is investigated both in
the one-body (OBDM) and in the two-body density matrix (TBDM). The
occurrence of the quantum phase transition is signaled by the vanishing
atomic condensate fraction and by the appearance of a gap in the
single-particle excitation spectrum.

Also other bosonic systems on a lattice, such as two-components BEC's, have
been shown to feature pair superfluidity~\cite{Kuklov04}, while our study is the first
to address this exotic quantum phase in continuum space using exact QMC techniques.
It is worth stressing that the pair superfluid phase considered here is found to be stable
against formation of cluster states. This requirement is not easily fulfilled by other
proposals of Bose condensates coupled by attractive forces where particles, due to
their statistics, are prone to collapse into bound states involving more than just a
dimer~\cite{Nozieres82}.

The microscopic Hamiltonian is defined as follows
\begin{eqnarray}
H&=&-\frac{\hbar^2}{2m}\sum_{i=1}^{N/2}\nabla_i^2-\frac{\hbar^2}{2m}
\sum_{\alpha=1}^{N/2}\nabla_\alpha^2
\nonumber\\
&+&
\sum_{i<j}\frac{d^2}{r_{ij}^3}+\sum_{\alpha<\beta}\frac{d^2}{r_{\alpha\beta}^3}
+ \sum_{i,\alpha}\frac{d^2(r_{i\alpha}^2-2h^2)}{(r_{i\alpha}^2+h^2)^{5/2}}
\;.
\label{hamiltonian}
\end{eqnarray}
The first two terms correspond to the kinetic energy of particles of
mass $m$ residing in the top layer (labeled by the index $i$) and in
the bottom layer (labeled by the index $\alpha$). Each layer contains
the same number $N/2$ of particles, $N$ being the total number of dipoles.
The terms in the second row of Eq.~(\ref{hamiltonian}) correspond to
the intra-layer and inter-layer dipolar interactions involving particles with
dipole moment $d$ oriented perpendicularly to the two layers separated
by a distance $h$. Here, $r_{ij(\alpha\beta)}=|{\bf r}_{i(\alpha)}-{\bf r}_{j(\beta)}|$
denotes the in-plane distance between pairs of particles in the top (bottom) layer,
and $r_{\alpha i}=|{\bf r}_\alpha-{\bf r}_i|$ is the distance between the projections
onto any of the layers of the positions of the $\alpha$-th and $i$-th particle.

Simulations are carried out in a 2D square box of area $A=L^2$ with periodic boundary conditions. 
The number of particles used ranges from  $N=60$ up to $N=180$ dipoles. 
The total density $n=N/A$ and the length scale $r_0=md^2/\hbar^2$, arising from the dipole-dipole force, 
define the relevant dimensionless parameters of the system: the interaction
strength $nr_0^2$ and the reduced distance $h/r_0$ between layers.
An important aspect of the physics of this bilayer configuration is the
existence of a bound state in the two-body problem for any value of
the inter-layer distance $h$~\cite{Simon76,Armstong10,Klawunn10}.  In
the following we denote by $\epsilon_b$ the binding energy of the
dimer which we obtain by solving numerically the Schr\"odinger
equation
$\left(-\frac{\hbar^2}{m}\nabla^2+\frac{d^2(r^2-2h^2)}{(r^2+h^2)^{5/2}}
-\epsilon_b\right)\psi_b({\bf r})=0$ for the pair wave function.

DMC simulations provide an exact result, in the statistical sense, for the
energy of the ground state of the system~\cite{Kolorenc11}.  
Simulations are greatly
speeded up when a physically relevant guiding wave function is used for 
importance sampling.  Although different choices of the guiding wave
function are not expected to affect the energy, non-local estimators might
be biased.  
In order to confirm the independence of the
results on the particular form of the guiding model, two different guiding
wave functions, containing  {\em a priori} different physics, have been
considered.  One is chosen to be of the Jastrow form
\begin{equation}
\Psi_T({\bf r}_1,\dots,{\bf r}_N)=\prod_{i<j}f_1(r_{ij})
\prod_{\alpha<\beta}f_1(r_{\alpha\beta})
\prod_{i,\alpha}f_2(r_{i\alpha}) \; ,
\label{trialfunction}
\end{equation}
where the in-plane and inter-layer two-body correlation terms, $f_1$ and
$f_2$, are non-negative functions of the pair relative coordinate.
In-plane correlations are parameterized using for $f_1$
the same functional form as in Ref.~\cite{Astra07}, which accounts
both for the cusp condition at short distances and for the phonon
contribution at large separations. The inter-layer term $f_2$ is
taken as the solution of the two-body problem up to $r=R_0$, imposing the
condition $f_2'(r=R_0)=0$, where $0\leq R_0 \leq L/2$ is
a variational parameter to be optimized. For distances larger than $R_0$
we set $f_2(r)=1$. As a second choice we have adopted an alternative guiding
function which explicitly accounts for the formation of pairs~\cite{Cazorla09}
\begin{eqnarray}
\Psi_T^{\text{pair}}({\bf r}_1,\dots,{\bf r}_N)=\prod_{i<j}f_1(r_{ij})
\prod_{\alpha<\beta}f_1(r_{\alpha\beta}) \times
\nonumber\\
\times \left( \prod_{i=1}^{N/2}\sum_{\alpha=1}^{N/2} \tilde f_2(r_{i\alpha}) +
\prod_{\alpha=1}^{N/2}\sum_{i=1}^{N/2} \tilde f_2(r_{i\alpha}) \right)\;.
\label{pairtrialfunction}
\end{eqnarray}

In-plane correlations are described in the same way by
$\Psi_T$ and $\Psi_T^{\text{pair}}$, while in~(\ref{pairtrialfunction})
inter-layer ones are rearranged in such a way that (though
preserving Bose symmetry in each layer) emergence of a given pairing
between particles in the bottom and top layer is favored. This is
achieved setting $\tilde f_2(r) = e^{-ar^2/(1+b r)}$ and
using $a$ and $b$ as variational parameters which we optimize.

\begin{figure}
\begin{center}
\includegraphics[width=0.95\linewidth,angle=0]{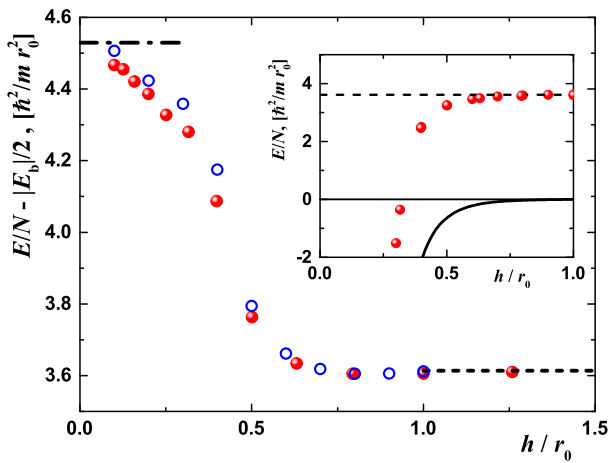}
\caption{(color online). Energy per particle with half of the dimer
  binding energy subtracted as a function of the reduced inter-layer
  distance $h/r_0$ for $nr_0^2=1$.
  Red solid and blue open symbols correspond to the results obtained using
  the guiding wave functions in Eq.~(\ref{trialfunction})
  and~(\ref{pairtrialfunction}), respectively.
  The horizontal lines correspond to
  the energies of a single layer of dipoles with effective interaction
  strength $\tilde{n}\tilde{r}_0^2=0.5$ and
  $\tilde{n}\tilde{r}_0^2=32$. Inset: Energy per particle ($nr_0^2=1$)
  and $\epsilon_b/2$ (solid line) as a function of the inter-layer
  separation. In both cases the dashed line corresponds to the large $h/r_0$
  limit.}
\label{fig2}
\end{center}
\end{figure}

The equation of state as a function of the inter-layer distance
$h/r_0$ at the density $nr_0^2=1$ is shown in Fig.~\ref{fig2}. 
The red solid and blue open symbols correspond to the results obtained using 
guiding wave functions~(\ref{trialfunction}) and~(\ref{pairtrialfunction})
extrapolated to the thermodynamic limit. The inset in Fig.~\ref{fig2} displays
the energy per particle $E/N$ compared to half of the dimer binding
energy $\epsilon_b/2$, while in the main figure we show the variation
of the difference $E/N-\epsilon_b/2$ with the distance $h/r_0$. We notice that
$E/N$ becomes negative when the inter-layer distance gets small enough and
approaches the dimer binding energy in the limit $h\ll r_0$. An important remark
is that in this regime the energy difference $E/N-\epsilon_b/2$ is found to be positive,
indicating that dimers feel an effective repulsive interaction which stabilizes the pair
phase.

Pairing between dipoles is in fact a strong effect when $h\ll r_0$, forming tightly
bound dimers which behave as composite objects featuring twice the mass and
dipole moment as compared to single dipoles. The horizontal lines in Fig.~\ref{fig2}
correspond to the energies per particle of a single layer of dipolar bosons with
an effective interaction strength $\tilde{n}\tilde{r}_0^2$, as obtained using the
results of Refs.~\cite{Astra07, Astra07_1}, where $\tilde{n}=n/2$ and the dipolar
length takes the two values $\tilde{r}_0=r_0$ and $\tilde{r}_0=8r_0$. The first
value corresponds to the asymptotic regime $h\gg r_0$ of independent layers,
whereas the second value refers to the opposite regime, $h\ll r_0$, where the
system behaves as a single layer of particles having dipole moment $2d$ and
mass $2m$ as mentioned above.

After discussing the equation of state we analyze the OBDM and TBDM as a
function of the inter-layer distance and the nature of the transition between
single-particle and pair superfluidity. The OBDM within each layer is defined as
\begin{equation}
\rho_1(s)=\frac{2}{n}\langle\psi_{t(b)}^\dagger({\bf r}+{\bf s})\psi_{t(b)}({\bf
  r})\rangle \;,
\label{OBDM}
\end{equation}
where $\psi_{t(b)}^\dagger({\bf r})$, $\psi_{t(b)}({\bf r})$ correspond to the
creation/annihilation operators of a particle at the 2D coordinate ${\bf r}$ in the
top (bottom) layer. The relevant TBDM involves instead a pair of particles residing
in different layers~\cite{Astra05}
\begin{equation}
\rho_2(s)=\frac{2}{n}\int d{\bf r}^\prime \langle\psi_b^\dagger({\bf
  r}+{\bf s})\psi_t^\dagger({\bf r}^\prime+{\bf s})\psi_t({\bf
  r}^\prime)\psi_b({\bf r})\rangle \;.
\label{TBDM}
\end{equation}
Notice the different normalization of the two functions at $s=0$:
$\rho_1(0)=1$ and $\rho_2(0)=N/2$.

For homogeneous systems, ODLRO in the OBDM implies a finite value of
Eq.~(\ref{OBDM}) at large separations:
$\lim_{s\to\infty}\rho_1(s)=n_0$, where $n_0\le1$ is the fraction of
atoms in the condensate of each layer. Similarly, ODLRO in
the TBDM entails that: $\lim_{s\to0}\rho_2(s)=\alpha$. One should
notice that ODLRO at the level of the OBDM implies ODLRO also at the
level of the TBDM and, in this case, $\alpha=(N/2)n_0^2$, which
is macroscopically large. However, even if ODLRO is absent in the
OBDM ($n_0=0)$, it can still be present in the TBDM, and $\alpha\,\,(<1)$
is interpreted as the condensate fraction of pairs~\cite{Yang62}.
An intrinsic order parameter related to the TBDM can be defined
as~\cite{Kobe69}: $\lim_{s\to\infty}\rho_2(s)-(N/2)n_0^2=n_0^{\text{mol}}$.
The molecular condensate fraction $n_0^{\text{mol}}$ coincides with
the long-range behavior $\alpha$ of the TBDM when the atomic
condensate $n_0$ vanishes and one removes from it the largest
contribution, which scales as the total number of particles, when
$n_0\neq 0$.

The calculation of the OBDM and TBDM using the DMC method relies on an
extrapolation technique based on both DMC and variational Monte Carlo
(VMC) results in order to extract the expectation value of the
relevant operator on the ground state of the system~\cite{Kolorenc11}.
The estimate obtained in this way is unbiased  only if the
guiding wave function $\psi_T$ used in the VMC calculation, and as
importance sampling  in the DMC simulation, do not differ too much
from the true ground-state wave function. In order to make sure that
our results for the density matrices do not depend on the special
choice of the guiding wave function (\ref{trialfunction}), we carried out DMC
simulations starting from the two wave functions in Eqs.~(\ref{trialfunction})
and~(\ref{pairtrialfunction}).

\begin{figure}
\begin{center}
\includegraphics[width=0.9\linewidth,angle=0]{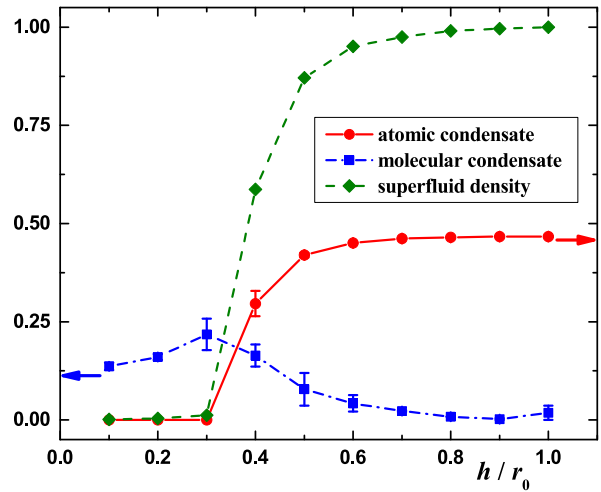}
\caption{(color online). Atomic condensate $n_0$ and molecular
  condensate $n_0^{\text{mol}}$ as a function of $h/r_0$ at the
  density $nr_0^2=1$. Arrows correspond to the condensate fraction of
  a single layer of dipoles at the effective interaction strength
  $\tilde{n}\tilde{r}_0^2=0.5$ (red arrow) and
  $\tilde{n}\tilde{r}_0^2=32$ (blue arrow).
  The dashed green line and symbols correspond to the super-counterfluid
  density of Eq.~(\ref{superfluid}).}
\label{fig3}
\end{center}
\end{figure}

Results for the OBDM and TBDM are shown in Fig.~\ref{fig3} for the
density $nr_0^2=1$. For both the particle and the molecular condensate, an
average between the determination using guiding function
(\ref{trialfunction}) and (\ref{pairtrialfunction}) has been performed,
and the error bars reported in the figures give an indication of how
close the two estimates are. The particle condensate is clearly
vanishing for inter-layer distances smaller than a critical value, and grows
continuously until it reaches the value corresponding to a single layer of dipoles
at the density $nr_0^2/2$~\cite{Astra07}. The molecular condensate fraction
$n_0^{\text{mol}}$ is extremely small in the regime of weak pairing corresponding
to large inter-layer separations, and increases smoothly in the region of the
transition to the molecular regime until it reaches the value expected for a single
layer of dipolar dimers. We notice that for values of $h$ in the range
$0.3\lesssim h/r_0\lesssim0.6$ both $n_0$ and $n_0^{\text{mol}}$ are appreciably
different from zero.

In Fig.~\ref{fig3} we also show the results of the super-counterfluid density
defined as~\cite{Pupillo13}
\begin{equation}
\rho_s=\lim_{\tau\to\infty}\frac{\langle ({\bf W}_t(\tau)-{\bf W}_b(\tau))^2\rangle}{2 N\tau}
\;,
\label{superfluid}
\end{equation}
and written in terms of the winding number relative to the top and bottom layer
\begin{equation}
{\bf W}_{t(b)}(\tau)=\sum_{i(\alpha)=1}^{N/2}\int_0^\tau d\tau^\prime
\left(\frac{d{\bf r}_{i(\alpha)}(\tau^\prime)}{d\tau^\prime} \right) \;,
\label{winding}
\end{equation}
where ${\bf r}_{i(\alpha)}$ is the in-plane coordinate of the particles
belonging to the top (bottom) layer. 
We notice that when the imaginary-time evolution of the particles in the top and bottom 
layer is 
fully correlated, as happens in the paired phase,
the contribution to $\rho_s$ vanishes, while $\rho_s=1$ when the two layers behave as independent superfluids. The results for $\rho_s$ and $n_0$ in Fig.~\ref{fig3} show that
both quantities vanish at the same critical inter-layer distance where the system enters the
pair superfluid phase. 
This behavior is consistent with a second-order phase transition, 
similarly to that reported in Ref.~\cite{Pupillo13} for the case of a bilayer system of dipolar bosons in an optical lattice.

\begin{figure}
\begin{center}
\includegraphics[width=0.9\linewidth,angle=0]{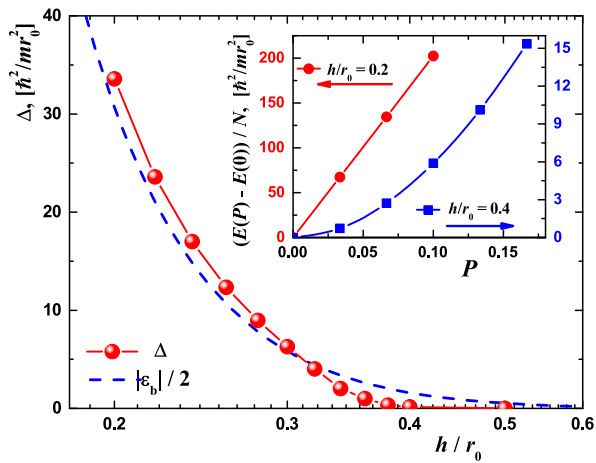}
\caption{(color online). Excitation gap $\Delta$ as a function of $h/r_0$ at the density $nr_0^2=1$. Red symbols and line: DMC results and guide to the eye. The dashed blue line corresponds to $\epsilon_b/2$. Inset: Energy $E(P)$ for two values of $h/r_0$ in the pair and single-particle
superfluid phase with the corresponding linear and quadratic fit.}
\label{fig4}
\end{center}
\end{figure}

\begin{figure}
\begin{center}
\includegraphics[width=0.9\linewidth,angle=0]{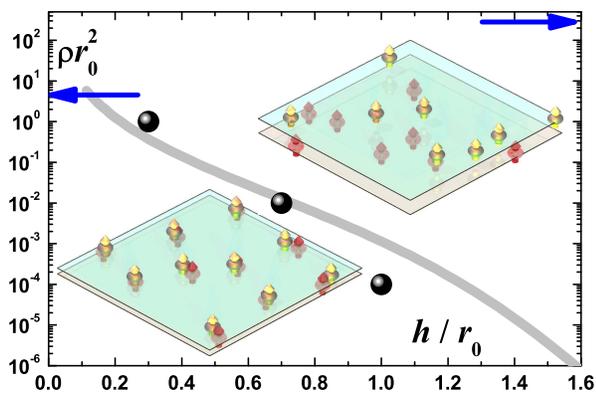}
\caption{(color online). Schematic phase diagram featuring the
  single-particle (upper region) and the pair superfluid (lower
  region). The dots correspond to the transition points as obtained from DMC
  simulations. The two arrows show the freezing density of a single layer of
  particles (right) and of dimers (left). The line  
  separates the region where $|\epsilon_b|/2<\mu$ (weak pairing) 
  from the region where $|\epsilon_b|/2>\mu$ (strong pairing)}
\label{fig1}
\end{center}
\end{figure}

The phase transition from the atomic to the pair superfluid is further characterized by the
appearance of a gap in the spectrum of elementary excitations. 
The gap value can be calculated from the dependence of the ground-state energy $E(P)$ on a small polarization
$P=(N_t-N_b)/N$ ($|P|\ll1$) obtained by slightly unbalancing the populations $N_{t(b)}$
of the top (bottom) layer while keeping the total number $N=N_t+N_b$ fixed. If an atomic condensate is present ($n_0\neq0$), the low-lying excitations are coupled phonon modes of
the two layers. In this case $E(P)=E(0)+N(n/2\chi_s)P^2$, where $E(0)$ is the ground-state energy
of the balanced system and $\chi_s$ is the spin susceptibility associated with the dispersion of
spin waves of the magnetization density $n_t-n_d$ with speed of sound $c_s=\sqrt{n/m\chi_s}$.
In the pair superfluid phase an energy $\Delta$ is needed to break a pair and spin excitations
are gaped. The resulting energy $E(P)=E(0)+N\Delta P$ is linear in the polarization. 
Examples of the 
different behavior of $E(P)$ is shown in the inset of Fig.~\ref{fig4} for two values of the
inter-layer distance corresponding to the atomic and pair superfluid. The values of the gap
$\Delta$ extracted from the fits to $E(P)$ are shown in Fig.~\ref{fig4}. The gap grows
continuously from zero starting from a critical inter-layer distance $h_c$ and for $h<h_c$
it approaches the value $\epsilon_b/2$ set by the dimer binding energy. The estimated value
$h_c\simeq0.35r_0$ for the critical distance is consistent with the results of the condensate and
super-counterfluid density of Fig.~\ref{fig3}.

Figures~\ref{fig2}-\ref{fig4} report results obtained for the gas parameter $nr_0^2=1$. 
Qualitatively similar results are found for different values of $nr_0^2$. 
A schematic phase diagram is shown in Fig.~\ref{fig1}, where the estimated critical inter-layer
distance $h_c$ is reported for three different densities. The figure also shows the
freezing density where the triangular solid is formed, occurring at
$\tilde{n}\tilde{r}_0^2\simeq290$~\cite{Astra07, Pupillo07}, for the
single layer of particles ($\tilde{r}_0=r_0$) and of pairs
($\tilde{r}_0=8r_0$).

In summary, in this work we have analyzed the behavior of a bilayer system
of perpendicularly aligned dipolar bosons in the continuum.  While the
two-body problem always has a bound state, we have shown that the many-body
system undergoes a phase transition from a molecular phase of tightly bound
pairs to a single-particle phase as the distance between the layers
increases.  Our simulations show that the phase transition is characterized
by the atomic condensate  fraction and the super-counterfluid density which
decay to zero simultaneously when the  inter-layer distance is reduced
below a certain critical value  that depends on the density.  At the same
critical value, a gap in the excitation spectrum opens and the system
enters the pair superfluid phase.  Remarkably, a bilayer system of dipoles
is, to the best of our knowledge, the first example of physical system
exhibiting bosonic pair superfluidity in the continuum that could be
explored in current experiments on ultracold gases with predominant dipolar
interactions.

We acknowledge partial financial support from the 
DGI (Spain) Grant No. FIS2011-25275 and  the Generalitat de Catalunya Grant 
No. 2009SGR-1003. 
GEA acknowledges a fellowship by MEC (Spain) through the Ramon y Cajal program.
SG acknowledges support from ERC through the QGBE grant and from  
Provincia Autonoma di Trento.

\end{document}